\documentclass[12pt]{iopart}        %preprint

\usepackage{iopams}

\expandafter\let\csname equation*\endcsname\relax

\expandafter\let\csname endequation*\endcsname\relax

\usepackage{amsmath}
\usepackage{bm}         % bold math
\usepackage{graphicx}   % Include figure files
\graphicspath{{./figures/}}

\usepackage{amsmath}
\usepackage{multirow}
\usepackage{cite}

\newcommand{\V}[1]{\bm{#1}} % alternate vector notation
\newcommand{\T}[1]{\bm{#1}} % rank 2 tensor notation
\newcommand{\ee}[1]{\mathrm{e}^{#1}} % exponential function

\newcommand{\dd}{\mathrm{d}} % typset 'd' for use in integral (e.g., dx)

\newcommand{\tens}{\mathop{\otimes}}

\newcommand{\avg}[1]{\left\langle#1\right\rangle}
\DeclareMathOperator{\Erfcx}{\mathrm{Erfcx}}
\DeclareMathOperator{\Erf}{\mathrm{Erf}}
\DeclareMathOperator{\Dsn}{\mathrm{D}}

\begin{document}
	\title{Nonlinear One-Dimensional Constitutive Model for Magnetostrictive Materials}% Force line breaks with \\
	\author{Alecsander N. Imhof}
	\ead{alecsanderi@vt.edu}
	\address{ Department of Biomedical Engineering and Mechanics, Virginia Tech.}
	\author{John P. Domann}
	\ead{jpdomann@vt.edu} %corresponding author email address
	\address{ Department of Biomedical Engineering and Mechanics, Virginia Tech.}
	%  \altaffiliation[Also at ]{Department of ...}
	
	\date{\today}% It is always \today, today,
	%  but any date may be explicitly specified
	
	\begin{abstract}
		This paper presents an analytic model of one dimensional magnetostriction. We show how specific assumptions regarding the symmetry of key micromagnetic energies (magnetocrystalline, magnetoelastic, and Zeeman) reduce a general three-dimensional statistical mechanics model to a one-dimensional form with an exact solution. We additionally provide a useful form of the analytic equations to help ensure numerical accuracy. Numerical results show that the model maintains accuracy over a large range of applied magnetic fields and stress conditions extending well outside those produced in standard laboratory conditions. A comparison to experimental data is performed for several magnetostrictive materials. The model is shown to accurately predict the behavior of Terfenol-D, while two compositions of Galfenol are modeled with varying accuracy. To conclude we discuss what conditions facilitate the description of materials with cubic crystalline anisotropy as transversely isotropic, to achieve peak model performance.
	\end{abstract}

	\newpage
	
	\section{Introduction}
	Magnetostrictive materials enable the use of numerous technologies including energy harvesters, ultrasonic transducers, vibration dampers, and even novel antennas \cite{Claeyssen1997, Davino2011, Deng2018, Lafont2012, Li2021, Narita2018a, Wang2008, Wang2016, Zenkour2020, Domann2017}. These materials intrinsically couple a material's magnetic and mechanical degrees of freedom, allowing the magnetization $\V{M}(\V{H},\T{\sigma})$ and magnetostriction $\T{\varepsilon_m}(\V{H},\T{\sigma})$ to be described as a function of the magnetic field $\V{H}$ and stress $\T{\sigma}$ (or total strain $\T{\varepsilon}$). However, it is difficult to accurately model the strongly coupled nonlinear magneto-mechanical constitutive response of these materials  at the macroscale. The lack of an accurate non-linear magnetostrictive constitutive model has likely inhibited the design of novel magnetostrictive technologies.
	
\begin{figure}[ht!]
		\centering
		\includegraphics[width = 8.5cm]{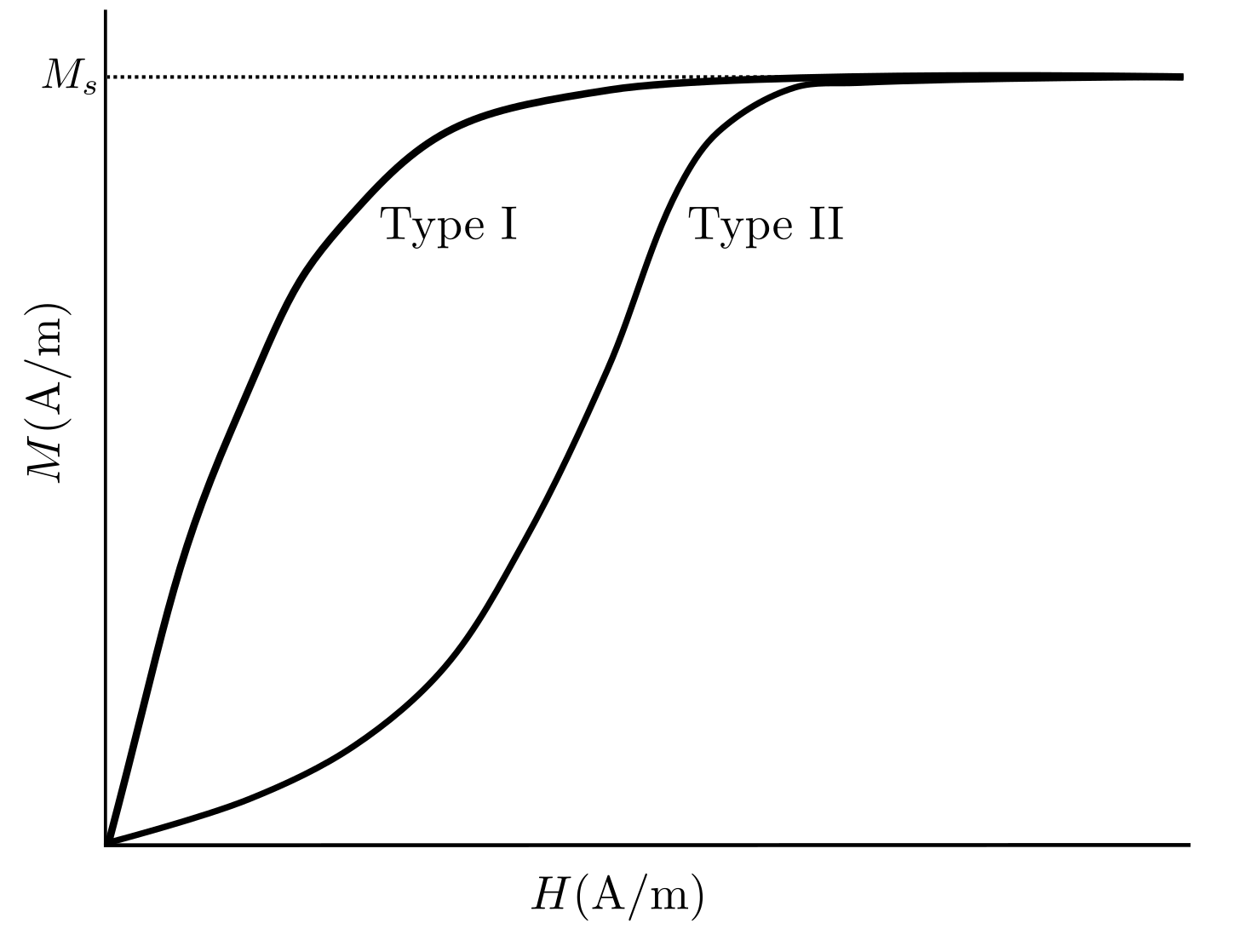}
		\caption{Schematic of two representative MH curves for magnetostrictive materials at fixed stress. Depending on the crystal structure of the material the responses can generally be  described by curves that are always concave down (Type I) or curves that transition from concave up to down. (Type II)}
		\label{fig:M-H}
	\end{figure}

	Figure \ref{fig:M-H} schematically illustrates two potential magnetization curves $\V{M}(\V{H},\T{\sigma}_0)$ at fixed stress $\T{\sigma}_0$. Two commonly encountered behaviors are highlighted. Type I curves are concave down until saturation, while Type II curves transition from concave up to concave down between demagnetization and saturation. Type I curves are commonly seen in materials like Terfenol-D ($\textrm{Tb}_{0.3}\textrm{Dy}_{0.7}\textrm{Fe}_{19.2}$) \cite{Zhao1998, Mahadevan2010}, while Type II curves are encountered in some compositions of Galfenol (e.g., $\textrm{Fe}_{81.6}\textrm{Ga}_{18.4}$) when placed in compression \cite{Mahadevan2010}. The specific MH curve displayed by a given material is strongly dependent on its crystalline structure in addition to any externally controlled anisotropies, like an applied stress. Numerous modeling approaches have been utilized in an attempt to capture the behaviors seen in Figure \ref{fig:M-H}. Three common approaches are to either 1) construct polynomial series expansions, 2) assume a modified Langevin behavior, or 3) utilize statistical mechanics \cite{Carman1995, Wan2003, Shi2016, Zhang2015, Zhou2016, Kim2020, Atulasimha2006, Evans2008a, Armstrong1997}.
	
	Polynomial series expansions start by constructing phenomenological Taylor series expansion of either the Helmholtz $f(\V{H}, \T{\varepsilon})$ or Gibbs $g(\V{H}, \T{\sigma})$ free energy density. Once a series is constructed $g \approx - \mu_0 \chi_{ij}H_i H_j /2 + ...$, classical thermodynamics is utilized to obtain conjugate field variables like the magnetization $\mu_0 \V{M} = -(\partial g/\partial\V{H})$ and magnetostriction $ \T{\varepsilon}_m = -(\partial g/\partial\V{\sigma})$. These models are straightforward to utilize once the expansion coefficients (i.e., material properties) are identified \cite{Carman1995, Wan2003}. However, conducting the requisite number of experiments to measure the expansion coefficients can be technically challenging and costly. Routinely only first order expansions are utilized, resulting in a preponderance of 'piezomagnetic' models, even though piezomagnetism is a fairly rare phenomenon most prevalent in antiferromagnets \cite{Dzialoshinskii1957, Newnham2005, Ibrahim2015, Yan2018}. One important restriction on polynomial approaches is immediately evident from the curves in Figure \ref{fig:M-H}. Notably, finite expansions are incapable of capturing the saturating behavior inherent in magnetic phenomena. As a result the use of polynomial models must be accompanied by knowledge of their bias conditions and limited range of validity. While the use of higher-order expansions can increase the range of validity, their use comes at the large cost of rapidly increasing the number of expansion coefficients. In practice unknowns or variations in test setups can result in difficulties using these linearized material properties when modeling an actual nonlinear device. 
	
	A common alternative to polynomial models is to \textit{a priori} assume a magnetic constitutive response that saturates and also assume micromagnetic expressions for magnetostriction are valid at the macroscale. We refer to this below as `Langevin magnetostriction'. An example of this approach is to assume $\T{\varepsilon_m} \propto \V{M}\tens{}\V{M}$, with $\V{M} = M_s \mathcal{L}(\V{h}) \V{\hat{h}}$ where $M_s$ is the saturation magnetization and $\mathcal{L}$ is the Langevin function depending on the reduced magnetic field $\V{h}$ \cite{Sablik1987,Li2011}. While the reduced field $\V{h}(\V{H}, T)$ is conventionally obtained from a ratio of the magnetic field energy divided by the thermal energy, several authors have made \textit{ad-hoc} assumptions that extend $\V{h}(\V{H}, \T{\sigma}, T)$ to depend on stress $\T{\sigma}$ as well \cite{Shi2016, Zhang2015, Zhou2016, Kim2020, Li2011, Sablik1987, Wang2011}. Langevin magnetostriction can qualitatively simulate a family of the Type I magnetization curves shown in Figure \ref{fig:M-H}, where stress modulates the susceptibility. While Langevin magnetostriction provides a path to incorporating saturating behavior and magneto-mechanical coupling, it also introduces inconsistencies compared to experimental observations, and more advanced models (e.g., micromagnetics). One large inconsistency introduced by Langevin magnetostriction concerns the $\Delta$E effect.  
	
	While the assumption $\T{\varepsilon_m} \propto \V{M}\tens{} \V{M}$ is often justified by appealing to the micromagnetic response $\T{\varepsilon_m} \propto \V{M} \tens{} \V{M}  /M_s^2  = \V{m}\tens{} \V{m} $, it is not generally possible to simply extend micromagnetic behavior to the macroscale without a suitable averaging procedure (e.g., it's more accurate to say $\T{\varepsilon_m} \propto \avg{ \V{m}\tens{} \V{m} } $, where $\avg{.}$ is a relevant thermal average). Consider that when $\V{H}=0$ an anhysteretic constitutive model predicts $\V{M}=0$, and therefore the assumptions above identically produce zero macroscopic magnetostriction $\T{\varepsilon_m}(\V{H}=0, \T{\sigma})= 0$. However, magnetic domains are easily controlled with stress at zero magnetic field, which is exemplified in numerous experimental studies that observe the peak $\Delta E$ effect occurs when $\V{H}=0$ \cite{Datta2010a, Datta2009a, Hubert2010}. This observed response is explicitly due to zero-field magnetostriction. Attempting to use Langevin Magnetostriction models while designing a device reliant on the $\Delta E$ effect, like a magnetometer \cite{Nan2008, Hatipoglu2015, Viehland2018}, is therefore expected to be a challenging endeavour. Instead of \textit{a priori} assuming $\V{M} = M_s \mathcal{L}(\V{h})\V{\hat{h}}$, we highlight that the Langevin model for classical paramagnetism is derived using statistical mechanics, indicating these assumptions likely do not need to be made.
	
	Models constructed with statistical mechanics are capable of capturing the full nonlinear saturating behavior and strong magneto-mechanical coupling of magnetostrictive materials. The Gibbs Canonical Ensemble can be used to model a magnetostrictive material in thermal equilibrium with a heat bath (the environment) at constant temperature, capable of having work done on it. As shown in Equations \eqref{eq:general_partition} - \eqref{eq: General Macroscopic Free Energy} this can be used to calculate the partition function $Z$ and the expected macroscale free energy $G$ for a given (microscopic) Landau free energy $G_L$ \cite{Bertotti1998}. 
	\begin{align}
		Z &= \int_{\mathbb{M}} \exp{\left(- \frac{G_L}{k_b T} \right)} \dd \V{m}    \label{eq:general_partition} \\
		G &= - k_b T \ln{Z} \label{eq: General Macroscopic Free Energy}
	\end{align}
	In these equations $k_b T$ is the thermal energy and integration is over all possible states of the system (i.e. $\mathbb{M}$ is the set of all admissible magnetization distributions). The Landau free energy contains any of the relevant energy densities used in micromagnetics. This allows the impact of external magnetic fields, magnetostatic fields (demag), magnetocrystalline anisotropy (MCA), exchange energy, and magnetostriction to be incorporated  \cite{Bertotti1998, Chikazumi2009, Cullity2009, OHandley2000, Kardar2007, Landau1976}. Once the expected energy $G$ is constructed, classical thermodynamics can be utilized and a series of partial derivatives leads to the average properties of the system. For a uniform system with volume $V$, the average magnetic moment $\avg{ \V{\mu} } =  \avg{ \V{M} } V =  -(\partial G / \partial \V{H} ) /\mu_0 $, and magnetostriction $\avg{ \T{\varepsilon}_m } =  -(\partial G / \partial \T{\sigma})/V$.
	
	While statistical  models can potentially model paramagnetic, ferromagnetic, ferrimagnetic, and antiferromagnetic materials with arbitrary anisotropy energies, and even account for polycrystalline materials \cite{Kardar2007,Landau1976,Atulasimha2008}, the requisite integral equations generally lack closed form solutions and therefore necessitate numerical approximations \cite{Atulasimha2006, Evans2008a, Armstrong1997}. This has resulted in common simplifying assumptions including zero exchange coupling (i.e., paramagnetic behavior) \cite{Smith2003, Atulasimha2008, Evans2010,Wahi2019}, potentially with simplified magnetocrystalline anisotropies that treat polycrystalline cubic materials as transversely isotropic \cite{Evans2009a, Evans2010, Evans2013, Wahi2019}.  Improving the computational efficiency of these models has been the focus of recent research that has shown an excellent ability to fit these models to experimental data \cite{Evans2008a, Evans2009a, Evans2010, Evans2013, Atulasimha2008, Atulasimha2011}.
	
	While generating a general 3D model is certainly a goal for this line of research, simplifying the model to one dimension has several benefits. Most notably, as we will show below there is a closed form analytical solution for a suitably simplified 1D model. Additionally, we note that most macroscale experimental studies have utilized conditions where the applied magnetic field and surface tractions are parallel $\V{H} \parallel \T{T}$ \cite{Atulasimha2011, Clark2000, Elhajjar2018, Mahadevan2010, Wun-Fogle2006, Moffett1991, Zhang2015, Zhao1998}. Each of the cited experimental studies has therefore provided effectively one-dimensional data, and not tested general 3D loading conditions. In addition to explaining existing experimental data, a valid 1D model is also expected to find use in reduced order models, including magnetostrictive rod and beam theories \cite{Nayfeh2005, Wang2017, Younis2003}. 
	
	The work in this paper presents an analytic constitutive model for magnetostrictive materials that is derived using statistical mechanics. This is made possible by restricting the allowed orientations of the applied magnetic field, stress, and MCA, in addition to restricting the allowable type of MCA. In the following sections we present the necessary assumptions, derive a 1D magnetostrictive constitutive model, provide a convenient numerical implementation, and use the model to simulate experimental data from the literature. Results show this model is capable of accurately simulating the response of materials with isotropic Joulian magnetostriction when the magnetocrystalline anisotropy of the material produces Type I magnetization curves as depicted in Figure \ref{fig:M-H}). 
	
	\section{Model Development}
	
	\subsection{Boltzmann Statistics}
	In this section we use the Gibbs Canonical Ensemble to describe a paramagnetic magnetostrictive material in thermal equilibrium with an environment at constant temperature, capable of having work done on it. For a paramagnetic material we focus on the average response of an isolated magnetic moment, allowing us to write the partition function and expected free energy density as
	\begin{align}
		z &= \int_{S} \exp{\left(- \beta g_L \right)} \dd \V{m}   \label{eq:local_partition} \\
		g &= - \beta^{-1} \ln{z} \label{eq:Macroscopic Free Energy}
	\end{align}
	where $g_L$ is the Landau free energy density, $\beta^{-1}$ the thermal energy density, and integration is now restricted to all orientations of $\V{m} = \V{M} / M_s$ (i.e., over the unit sphere $S$). Shown in Equation \eqref{eq:total_energy}, we consider a free energy density composed of Zeeman $f_z$, magnetoelastic anisotropy $f_{me}$, and magnetocrystalline anisotropy $f_{mca}$ energy densities \cite{Bertotti1998, Chikazumi2009, Cullity2009, OHandley2000}. 
	\begin{align}
		g_L &= f_z(\V{m};\V{H}) + f_{me}(\V{m};\T{\sigma}) + f_{mca}(\V{m};\T{K}) \label{eq:total_energy}
	\end{align}
	where the notation $f(\V{m};...)$ indicates $\V{m}$ is a prescribed parameter and $\T{K}$ is the MCA tensor. Once the partition function is constructed the equilibrium magnetization and magnetostriction are calculated by taking partial derivatives of the expected free energy in Equation \eqref{eq:Macroscopic Free Energy}, leading to
	\begin{align}
		\avg{ \V{M} } 
		&= M_s \frac{1}{z} \int_{S} \V{m} \exp{\left(- \beta g_L \right)} \dd \V{m} \nonumber \\
		&= M_s \avg{\V{m}} \label{eq:magnetization}\\
		\avg{\T{\varepsilon}_m} 
		&= \frac{3 \lambda_s}{2}  \frac{1}{z} \int_{S} \V{m} \tens{} \V{m} \exp{\left(- \beta g_L \right)} \dd \V{m}  \nonumber\\
		&= \frac{3}{2} \lambda_s \avg{ \V{m} \tens \V{m} } \label{eq:magnetostriction}
	\end{align}
	where $\lambda_{s}$ is the saturation magnetostriction. Based on Equations \eqref{eq:magnetization} and \eqref{eq:magnetostriction} for the average magnetization and magnetostriction, we can identify the probability density function $P(\V{m}; \V{H}, \T{\sigma}, \beta) = \exp{(-\beta g_L)} / z$.  This allows equations \eqref{eq:magnetization} and \eqref{eq:magnetostriction} to be interpreted as thermal averages of the micromagnetic equations for magnetization and isotropic Joulian magnetostriction. 
	
	In addition to obtaining the magnetization and magnetostriction, the nonlinear material properties can be calculated by taking derivatives of the previous expressions. The properties are readily shown to be
	\begin{align}
		\avg{\T{\chi}} 
		&= \beta \mu_0 (\avg{\V{M} \tens \V{M}} - \avg{\V{M}}\tens \avg{\V{M}})
		\label{eq:chi_general}\\
		\avg{\T{S}_{\V{m}}} &= \beta (\avg{\T{\varepsilon}_m \tens \T{\varepsilon}_m} 
		- \avg{\T{\varepsilon}_m}\tens \avg{\T{\varepsilon}_m}) 
		\label{eq:S_general}\\
		\avg{\T{q}} &= \beta (\avg{\V{M} \tens \T{\varepsilon}_m} 
		- \avg{\V{M}}\tens \avg{\T{\varepsilon}_m})  
		\label{eq:q_general}
	\end{align}
	where $\avg{\T{\chi}}$ is the magnetic susceptibility at constant stress, $\avg{\T{S}_{\V{m}}}$ the magnetostrictive compliance at constant magnetic field, and $\avg{\T{q}}$ the piezomagnetic coupling tensor defined by $\avg{\T{q}} = \partial \avg{\V{M}} / \partial{\T{\sigma}} = \mu_0^{-1} \partial \avg{\T{ \varepsilon_m}} / \partial \V{H}$. We briefly note that the total compliance $\T{S} = \T{S}_{el} + \avg{\T{S}_{m}}$, where $\T{S}_{el}$ is the elastic compliance (e.g., inverse Young's modulus for 1D loading). The form of Equations \eqref{eq:chi_general}-\eqref{eq:q_general} reveals that the macroscopic material properties are proportional to the statistical variance / fluctuation of the underlying microscopic fields (i.e., $\mathrm{var}(x) = \avg{x^2} - \avg{x}^2$). 
	
	We briefly note that equations \eqref{eq:local_partition} - \eqref{eq:q_general} can commonly be simplified by shifting the free energy with functions that are independent of $\V{m}$. The magnetization $\langle\V{\tilde{M}}\rangle$ and magnetostriction $\langle\T{\tilde{\varepsilon}_m}\rangle$ obtained from the free energy $\tilde{g}_L = g_L + f(\T{H}, \T{\sigma})$ is related to the average magnetization $\avg{\V{M}}$ and magnetostriction $\langle\T{\varepsilon}_m\rangle$ obtained from $g_L$ by 
	\begin{align}
		\tilde{z} 
		&= \int_{S} \exp{\left(- \beta \tilde{g_L} \right)} \dd \V{m} = z \, \exp(-\beta f) 
		\label{eq:partition prime}\\
		\avg{ \V{\tilde{M}} } 
		&= -\frac{1}{\mu_0}  \frac{\partial f}{\partial \V{H}} + \avg{ \V{M} } 
		\label{eq:magnetization prime}\\
		\avg{ \T{\tilde{\varepsilon}}_m}
		&= -  \frac{\partial f}{\partial\T{\sigma}} + \avg{ \T{\varepsilon}_m } 
		\label{eq:magnetostriction prime} .
	\end{align}
	In practice, a function independent of $\V{m}$ can be added to $g_L$ to ensure the Boltzmann term $\exp{(-\beta \tilde{g}_L)} \leq 1$ which aids in numerical calculations (i.e., so the exponential doesn't lead to numerical inaccuracies when $g_L << -1$). In what follows, terms with tilde overbars are understood to include an energy offset as described in Equations \eqref{eq:partition prime}-\eqref{eq:magnetostriction prime}.
	
	As previously stated the integrals used in Boltzmann statistics do not generally possess closed form solutions. However, we now show that exact closed form solutions exist for a specific type of MCA and restrictions on the orientations of the magnetic field and stress.
	
	\subsection{Quadratic Anisotropy}
	\label{subsection: Quadratic Anisotropy}
	
	A closed form solution to Equations \eqref{eq:local_partition} - \eqref{eq:q_general} can be obtained under the following assumptions: 1) the MCA can be represented as a quadratic form $f_{mca} = \V{m}\cdot \T{K} \V{m}$, 2) the material displays isotropic magnetostriction (i.e., $\lambda_{100} = \lambda_{111}= \lambda_s$), 3) the combined magnetoelastic anisotropy and MCA is transversely isotropic, and 4) the magnetic field is perpendicular to the isotropic anisotropy plane. To satisfy these requirements we assume the material has transversely isotropic MCA, with only two unique eigenvalues $K_i$. We assume the unique direction to be the 1-direction, while the 23-plane is isotropic (i.e., $K_1 \neq K_2 = K_3$). Additionally, we assume the stress state has only 2 unique principal stresses $\sigma_1 \neq \sigma_2 = \sigma_3$, where the stress and MCA eigenvectors are parallel. These stress assumptions are commonly satisfied for long rod or beam-like materials with one unique axis, and are consistent with numerous experimental studies \cite{Atulasimha2011, Clark2000, Elhajjar2018, Mahadevan2010, Wun-Fogle2006, Moffett1991, Zhang2015, Zhao1998}. The assumption of transversely isotropic MCA is a key mathematical assumption in this model that restricts the possible materials this model is suitable for. This point will be examined further in the results section.
	
	To see how these assumptions lead to a closed form solution we first construct the total anisotropy energy density $f_A$
	\begin{align}
		f_A &= f_{mca} + f_{me} \\
		f_{mca} &= \V{m}\cdot \T{K} \V{m} = K_{ij} m_i m_j\\
		f_{me} &= \V{m} \cdot \T{\Sigma} \V{m} = -\frac{3}{2} \lambda_{s} \sigma_{ij} m_i m_j, 
		\label{eq:f_me}
	\end{align}
	where the isotropic magnetostriction is described by $\T{\Sigma} = -3 \lambda_s \T{\sigma}/2$ and $\T{\sigma}$ is the Cauchy stress. Both $\T{K}$ and $\T{\Sigma}$ are symmetric rank 2 tensors. We combine them to define the non-dimensionalized anisotropy tensor $\T{A} = -\beta \left( \T{K} + \T{\Sigma} \right)$. The non-dimensionalized magnetic field is $\V{h} = \beta \mu_0 M_s \V{H}$. 
	
	Expressing these energies with components parallel to the eigenvectors of $\T{A}$,  $-\beta g_{L} = A_i m_i^2 + h_i m_i$, where $A_i$ are the eigenvalues of $\T{A}$. Additionally, the transversely isotropic $\T{A}$ has only two unique eigenvalues $A_1 \neq A_2 = A_3$, and following assumption 4) $\V{h} = h \V{\hat{e}_1}$ is parallel to the 1-axis. Accounting for the fact that $|\V{m}| = 1$ is a unit vector, and discarding the resultant term independent of $\V{m}$, we have $-\beta g_{L} = A m_1^2 + h m_1$, where $ A = A_1 - A_2$ is the change in anisotropy energy from the isotropic plane to the transverse axis. Summarizing, based on the assumptions above we can simplify Equation \eqref{eq:local_partition} to 
	\begin{align}
		z &= 2 \pi \int_{-1}^{1} \exp{\left(A m_1^2 + h m_1 \right)} \dd m_1
		\label{eq:1D Partition}
	\end{align}
	
	The solution to the integral in \eqref{eq:1D Partition} is a summation of Dawson functions $\Dsn()$. However, when $A<0$ the Dawson function produces complex numbers and calculations using this function can rapidly accumulate large numerical errors. In that case the integral can be simplified and written in terms of the error function $\Erf()$ as summarized in Equation\eqref{eq:partition function 1}  
	\begin{align}
		\frac{z}{2 \pi} &= \begin{cases} 
			z_{+} = \frac{\exp{\left( A + h \right)}\Dsn(\alpha_+) + \exp{\left(A - h \right)}\Dsn(\alpha_-)}{\sqrt{A}} 
			& A > 0 \\
			z_{-} = -\frac{\exp{\left(- \frac{h^2}{4A} \right)} \sqrt{\pi} (\Erf(\gamma_+) - \Erf(\gamma_-))}{2 \sqrt{-A}} 
			& A < 0 \\
		\end{cases} \label{eq:partition function 1}
	\end{align}
	where $\alpha_{\pm} = (2 A \pm h)/(2\sqrt{A})$ and $\gamma_{\pm} = (h \pm 2 A)/(2\sqrt{-A})$. Following equations \eqref{eq:magnetization} to \eqref{eq:q_general} above, derivatives of these expressions can be used to obtain the magnetization, magnetostriction, and nonlinear material properties. However, we first make several refinements to these equations as there are numerous points where, although they are analytically correct, they can accrue large numerical errors when evaluated.
	
	The first point at which the current solutions can become inaccurate is when $h>>|A|$. Notice that in \eqref{eq:partition function 1} as $A \rightarrow 0$ the equations become indeterminate. A simple fix for this problem is to utilize a series expansion of the partition function (Equation \eqref{eq:1D Partition}) for small anisotropy values about $A=0$. The general form of the series expansion is provided in the Appendix in section \ref{subsection: Appendix Series Expansion}. Additionally this approximation can be used to solve for the magnetization, magnetostriction, and material properties. The series expansion should be used when $|A|/h < \epsilon$, where $\epsilon$ will depend on the specific numerical implementation. When testing second order expansions using Matlab we found a cutoff ratio of $|A|/h < 10^{-7}$ preserved accuracy in the magnetization and magnetostriction.
	
	The use of the error function $\Erf(x)$ is susceptible to numerical errors when $x >> 1$. The numerical accuracy can be improved by introducing the scaled complimentary error function $\Erfcx(x)$, where $\Erf(x) = 1 - \exp(-x^2) \Erfcx(x)$. This has the advantage of simplifying several exponential terms, and helping to avoid arithmetic underflow \cite{Cody1993}. As previously stated these solutions can be further simplified by shifting the free energy. The specific energy offsets used below were all chosen to ensures that exponential terms in the solutions remain bounded as the magnetic field and stress become large. Combined these changes simplify Equation \eqref{eq:partition function 1} to 
	\begin{align}
		\frac{\tilde{z}}{2 \pi} &= \begin{cases} 
			\tilde{z}_{+} = \frac{\Dsn(\alpha_+) + \exp{\left(-2 h \right)} \Dsn(\alpha_-)} {\sqrt{A}} 
			& 
			A > 0 \\
			\tilde{z}_{-} = -\frac{\sqrt{\pi} (- \Erfcx(\gamma_{+}) + \exp{\left( -2 h  \right)} \Erfcx(\gamma_{-})) }{2 \sqrt{-A}}
			&  
			A < 0, \, \gamma_{+} > 0 \\
			\tilde{z}_{-} = -\frac{\sqrt{\pi} (-2 + \exp{\left( -\gamma_{+}^2 \right)} \Erfcx(\gamma_{+}) + \exp{\left( -\gamma_{-}^2 \right)} \Erfcx(\gamma_{-})) }{2 \sqrt{-A}}
			&
			A < 0, \, \gamma_{+} < 0, \\
		\end{cases} \label{eq:zprime}
	\end{align}
	where the energy offsets applied to Equation \eqref{eq:zprime} are
	\begin{align}
		f &= \begin{cases} 
			-A - h 
			& 
			A > 0 \\
			-A - h
			&  
			A < 0, \, \gamma_{+} > 0 \\
			h^{2}/4 A
			&
			A < 0, \, \gamma_{+} < 0. \\
		\end{cases} \label{eq:offsets}
	\end{align}
	These offsets were selected to ensure the exponential terms in \eqref{eq:partition function 1} converge to zero as $\{h, \, A\} \rightarrow \infty $. The derivatives of equations \eqref{eq:zprime} and \eqref{eq:offsets} with respect to $h$ results in the average magnetization,
	\begin{align}
		\frac{\avg{ M }}{M_s} &= \begin{cases} 
			-\frac{h}{2A} 
			- \frac{-1 + \exp(-2 h)}{2A \, \tilde{z}_{+}} 
			& 
			A > 0 \\
			-\frac{h}{2A} 
			- \frac{1 - \exp(-2 h)}{2A \, \tilde{z}_{-}}
			&  
			A < 0, \, \gamma_{+} > 0 \\
			-\frac{h}{2A} 
			+ \frac{\exp{\left(-\gamma_{+}^2 \right)} - \exp{\left( - \gamma_{-}^2 \right)} }{2A \, \tilde{z}_{-}}
			&
			A < 0, \gamma_{+} < 0. \\
		\end{cases} \label{eq:Magnetization 1}
	\end{align}
	Additionally, the derivative of  \eqref{eq:partition function 1} with respect to $A$ results in an average magnetostriction of
	\begin{align}
		\frac{2}{3} \frac{\avg{ \varepsilon_{m}} }{\lambda_{s}} &= \begin{cases} 
			\frac{h^2-2A}{4 A^2} 
			+ \frac{h}{\tilde{z}_{+}} \left(
			\frac{ 1 + \exp{\left(-2 h \right)}}{2A h} 
			- \frac{  1 - \exp{\left(-2 h \right)}}{4 A^2}
			\right)
			&
			A > 0 \\
			\frac{h^2-2A}{4 A^2} 
			+ \frac{h}{\tilde{z}_{-}} \left(
			\frac{ 1 + \exp{\left(-2 h \right)}}{2A h} 
			- \frac{ 1 - \exp{\left(-2 h \right)}}{4 A^2}
			\right)
			& 
			A < 0, \, \gamma_{+} > 0 \\
			\frac{h^2-2A}{4 A^2} 
			+ \frac{h}{\tilde{z}_{-}} \left(
			\frac{ \exp{\left(\gamma_{+}^2 \right)} + \exp{\left( - \gamma_{-}^2 \right)}}{2A h} 
			- \frac{ \exp{\left(\gamma_{+}^2 \right)} + \exp{\left( - \gamma_{-}^2 \right)}}{4 A^2}
			\right)
			& 
			A < 0  ,\, \gamma_{+} < 0 
		\end{cases} \label{eq:Magnetostriction 1}
	\end{align}
	Expressions for the nonlinear material properties in Equations \eqref{eq:chi_general}
	- \eqref{eq:q_general} can be obtained through additional derivatives of the equations above. As these expressions are reasonably lengthy, we provide them in section \ref{subsection:Appendix Material Properties} of the Appendix.
	
	As a final note, the equations above are written assuming that $h \geq 0$. Values for $h<0$ can readily be obtained by noting the magnetization is an odd function with respect to $h$, while the magnetostriction is even (i.e., for $h<0$ $M(h) = -M(|h|)$. We also note that in contrast to the Langevin Magnetostriction models discussed in the introduction, when $h = 0$ the magnetostriction $\avg{\varepsilon_m} \neq 0$ in this model. Instead, when $h=0$ a net magnetostriction is induced due to the competing MCA and magnetoelastic anisotropy energies. Finally, similar equations for $\avg{M}$ and $\avg{ \varepsilon_{m} }$ have previously appeared in the literature, most closely matching the results in Equation \eqref{eq:Magnetization 1} and \eqref{eq:Magnetostriction 1} when $A>0$ \cite{Raghunathan2009, Talleb2020}. However the authors believe this is the first time a numerically accurate solution for both $A<0$ and $A>0$ has been presented. 
	
	\section{Results and Discussion}
	This section provides 1) a numerical comparison of the closed form solutions presented in equations \eqref{eq:zprime}, \eqref{eq:Magnetization 1}, and \eqref{eq:Magnetostriction 1} to conventional numerical integration, 2) a qualitative assessment of material response these solutions produce, and finally 3) assesses the model's ability to simulate the experimentally measured response of several common magnetostrictive materials.
	
	\subsection{Numerical Accuracy}
	\label{subsection: Numerical Accuracy}
	
	To evaluate the numerical accuracy of these solutions they were compared to standard numerical integration. A grid of $N=100$ logarithmically spaced field points $10^{-2} \leq h\leq 10^{6}$ and $N=200$ logarithmically spaced anisotropies  $10^{-2} \leq |\pm  A| \leq 10^{6} $ were generated for $N_{total} = 20,000$ points. At each grid point numerical integration was performed using Matlab's \texttt{integral()} function with relative and absolute errors of $10^{-12}$.  The relative errors for each equation were calculated as $|f_{num} - f_{eqn}| / f_{num}$, where $f$ is the parameter of interest. While we utilized Matlab's built-in scaled complimentary error function \texttt{erfcx()}, the built-in Dawson function was quite slow, and instead we approximated $\Dsn()$ using McCabe's continued fraction. This achieves a precision to $10^{-15}$, and has a simple and fast numerical implementation \cite{McCabe1974}. 
	
	\begin{table}
		\caption{Accuracy compared to numerical integration \label{table:accuracy} } % title
		%\begin{indented}
		%\item[]
		\centering
		\begin{tabular}{@{}*{3}{c}} % centered columns (3 columns)
			\br % bold line 
			Function & Avg. Error & Max Error \\
			\mr % inserts line
			% Partition Function 
			$\tilde{z}$  & $2.4\times10^{-3}$ & $6.8\times10^{-1}$  \\
			% Magnetization
			$\avg{M}$ & $1.9\times10^{-11}$ & $2.3\times10^{-9}$ \\  
			% Magnetostriction
			$\avg{\varepsilon_{m}}$ & $2.4\times10^{-5}$ & $3.9\times10^{-3}$ \\ 
			\br
		\end{tabular}
		%\end{indented}
	\end{table}
	
	The accuracy results are summarized in Table \ref{table:accuracy}. Additionally, detailed error surfaces are shown in section \ref{subsection: Appendix Error Surfaces} of the Appendix. The data in Table \ref{table:accuracy} shows 1) the average relative error per grid point (i.e., sum of individual errors divided by $N_{total})$ and 2) the maximum relative error at a single grid point. The relative errors of all three functions were approximately $10^{-14}$ for the majority of the tested grid points. The only region where the solutions became inaccurate was when the series expansion was used for $|A|/h < 10^{-7}$. The maximum errors were all measured in the expansion region. It should be noted that without the series expansion the error can quickly climb above 100\% in that region (i.e., the expansion worked). While the partition function had a maximum error of over $50\% $ for the expansion, the resulting magnetization and magnetostriction still remained accurate as they depend on derivatives of $z$ not its absolute value. The max relative error in the magnetostriction only reached $0.4 \%$. 
	
	To provide more context for this information, if we consider the material properties for Terfenol-D that are presented in Section \ref{subsection: Exp Comp} below, we can show that accuracy analysis above is valid for applied fields $0 \leq H \leq 200$ T and stresses $0 \leq \sigma \leq 10$ TPa, respectively. This range is clearly far outside those used in standard laboratory settings, leaving the authors to conclude the presented equations remain numerically accurate for fields and stress conditions typically applied to these materials. Material failure and additional phenomena would need to be considered before these equations become numerically inaccurate.  
	
	\subsection{Qualitative Assessment}
	\begin{figure}[ht!]
		\centering
		\includegraphics[width = 15cm]{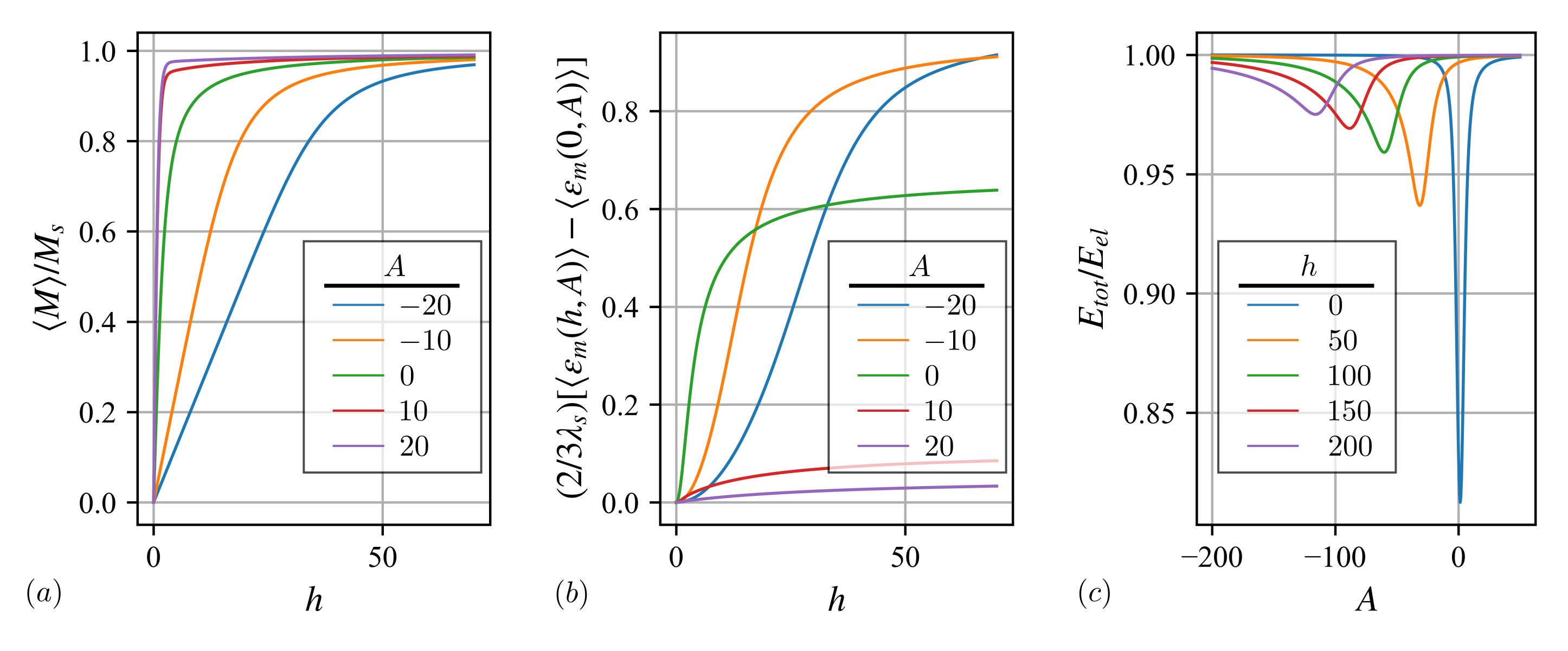}
		\caption{Model predictions of (a) magnetization and (b) magnetostriction and (c) the $\Delta E$ effect for possible $h$ and $A$ values.}
		\label{fig:QA}
	\end{figure}
	
	Having shown the equations above accurately solve the integral expressions used in statistical mechanics, we now turn our attention to analyzing the qualitative response of the resulting model. Figure \ref{fig:QA} illustrates the (a) magnetization and (b) magnetostriction curves at fixed anisotropy $A$. Figure \ref{fig:QA} (c) shows the $\Delta E$ effect this model predicts at fixed values of $h$.
	Starting with Figure \ref{fig:QA} (a) we note that in addition to properly saturating, the model has a lower susceptibility when A is negative (compression) and a higher susceptibility when A is positive (tension). Furthermore, when $A=0$ the model reduces to the Langevin function as required. Finally, while only small values of $A$ are presented in Figure \ref{fig:QA}(a), over the entire range of tested values $10^{-2} \leq |\pm A| \leq 10^{6}$ only Type I behavior was observed. This model does not qualitatively describe materials with Type II MH curves.
	
	Figure \ref{fig:QA} (b)  shows the predicted magnetostriction curves as a function of $h$ with different lines at fixed anisotropy $A$.
	Note that the zero field magnetostriction has been subtracted from each curve to clearly display all curves on the same graph. Therefore the plot shows $\avg{\varepsilon_m(h,A)} - \avg{\varepsilon_m(0,A)}$. For a positive magnetostrictive material in tension $A>0$, the model correctly predicts a small change in magnetostriction as $h$ increases. For this case tension produces an initial magnetization configuration where the microscale magnetization starts either parallel or anti-parallel to the magnetic field direction. Upon increasing $h$ the anti-parallel domains transition to parallel, however there is no resulting magnetostriction as $\varepsilon_m \propto m_1^2$ is even in the magnetization. Conversely, compression causes the magnetization to initially align perpendicular to the applied field direction. Increasing the applied magnetic field then forces the domains to become parallel. For a large enough initial compression this results in $\avg{\varepsilon_m(h,A)} - \avg{\varepsilon_m(0,A)} = 3 \lambda_s /2$ as $h$ increases.
	
	Concluding with Figure \ref{fig:QA} (c) the total Young's modulus of the material $E_{tot}$ was calculated using the using the magnetostrictive compliance $S_m$ and elastic Young's Modulus $E_{el}$, 
	\begin{align}
		E_{tot} = (1/E_{el} + \avg{S_m})^{-1}.
	\end{align} 
	As the intention of Figure \ref{fig:QA} is qualitative assessment of the predicted behavior, a Young's Modulus of $1$ Pa was chosen. Changing this value changes the amplitude of $E_{tot}$ and $\Delta E$, but not the locations of the peaks. The trends in this graph are consistent with experimental data in which the largest $\Delta E$ effect is observed at zero magnetic field \cite{Datta2010a}. Furthermore, recalling that A is composed of MCA and magnetoelastic anisotropy, this model shows that in order to maximize the $\Delta E$ effect an applied stress that cancels out MCA is required. The maximum value of $\avg{S_m}$ obtained from Equation \eqref{eq:S_general} is equal to $\beta \lambda_{s}^2 /5$, leading to the maximum $\Delta E = E_{el} - E_{tot} $.
	\begin{align}
		\max \Delta E = \left( \frac{1}{E_{el}} +  \frac{5}{\beta \lambda_{s}^2 E_{el}^2 } 
		\right)^{-1}
	\end{align}
	
	\subsection{Experimental Comparison \label{subsection: Exp Comp}}
	In addition to confirming the numerical accuracy of the magnetization and magnetostriction in Equations \eqref{eq:Magnetization 1} and \eqref{eq:Magnetostriction 1}, the solutions were compared to magnetization and magnetostriction curves found in the literature for three different materials. We digitized data for Terfenol-D $\mathrm{Tb}_{0.3} \mathrm{Dy}_{0.7} \mathrm{Fe}_{19.2}$ \cite{Zhao1998}, and two compositions of Galfenol $\mathrm{Fe}_{79.1} \mathrm{Ga}_{20.9}$, and $\mathrm{Fe}_{81.6} \mathrm{Ga}_{18.4}$ \cite{Mahadevan2010}. The saturation magnetization $M_s$ and saturation magnetostriction $\lambda_s$ were directly obtained from the experimental data, with $M_s = \max(|M|)$, and $\lambda_s = 2/3 \max(|\varepsilon_m|)$. The magnetostriction calculation came from curves where we assume the initial bias stress was large enough to exclusively produce $180^\circ$ domain walls in the material when $H=0$. After $M_s$ and $\lambda_s$ were obtained, the only remaining unknowns in the model are the magnetocrystalline anisotropy coefficient $K$, and the thermal energy density term $\beta$. For each material $K$ and $\beta$ were identified by minimizing the relative error between the constitutive model and the experimental data. This calculation produces error surfaces that only depend on two variables. Therefore we did not use a nonlinear optimization routine, but instead calculated the error over a grid of $K$ and $\beta$ values. These error surfaces are presented in the results below.  
	
	The values of $K$ and $\beta$ obtained from this procedure, along with the average and maximum relative error between the modeled and experimental data are presented in Table \ref{table:properties}. As we calculate errors when comparing the 1) magnetization and 2) magnetostriction data, we present three sets of values for $\{K,\beta\}$, and their errors for each material. The three cases are for 1) fitting the combined data set, 2) fitting only the magnetization data, and 3) fitting only the magnetostriction data. The last two fits can be utilized for models requiring only $\avg{M}$ or $\avg{\varepsilon_m}$ (i.e., in one-way-coupled models). 
	
	\begin{table}
		\caption{Model Parameters and Results* \label{table:properties}} % title of Table
		\begin{indented}
			\item[] 
			\begin{tabular}{*{8}{c}} % centered columns (8 columns)
				\br % bold line 
				\multirow{2}{*}{Material} & \multirow{2}{*}{Fit} & \multirow{2}{*}{$K$} & \multirow{2}{*}{$\beta$} & \multicolumn{2}{c}{Avg. Error} & \multicolumn{2}{c}{Max. Error}\\
				& & & & $\avg{M}$  & $\avg{\varepsilon_{m}}$ & $\avg{M}$ & $\avg{\varepsilon_{m}}$\\
				\mr % inserts single horizontal line
				% Terfenol Data 
				$\textrm{Tb}_{0.3}\textrm{Dy}_{0.7}\textrm{Fe}_{19.2}$
				& Combined & -4697 & $6.5\times10^{-4}$ & 7.8 \% &  3.0 \% & 33 \% & 20 \%\\ % Combined 
				$M_s$ = $7.8\times10^{5}$
				& Only $\avg{M}$ & 6515 & $1.6\times10^{-4}$ & 5.0 \% & --- & 55 \% & --- \\ % Magnetization 
				$\lambda_s$ = $1.4\times10^{3}$
				& Only $\avg{\varepsilon_{m}}$ &  -4697 & $6.6\times10^{-4}$ & --- & 3.0 \% & --- & 19 \%\\ %  Magnetostriction
				\mr
				% Galfenol 1 Data
				$\textrm{Fe}_{79.1}\textrm{Ga}_{20.9}$
				& Combined & 363.6 & $1.1\times10^{-3}$ & 6.4 \% & 12 \% & 26 \% & 38 \%\\ % Combined
				$M_s$ = $1.2\times10^{6}$
				& Only $\avg{M}$ &  -767.7 & $7.4\times10^{-4}$ & 4.0 \% & --- & 13 \% & ---\\ % Magnetization
				$\lambda_s$ = $1.3\times10^{2} $
				& Only $\avg{\varepsilon_{m}}$ & 1010 & $2.8\times10^{-3}$ & --- & 6.7 \% & --- & 37 \%\\ %  Magnetostriction
				\mr
				% Galfenol 2 Data
				$\textrm{Fe}_{81.6}\textrm{Ga}_{18.4}$
				& Combined & 909.2 & $1.7\times10^{-3}$ & 6.8 \% & 7.0 \% & 23 \% & 31 \%\\ % Combined 
				$M_s$ = $1.2\times10^{6}$
				& Only $\avg{M}$ & 1333 & $1.6\times10^{-3}$ & 6.5 \% & --- & 22 \% & ---\\ % Magnetization 
				$\lambda_s$ = $1.7\times10^{2}$
				& Only $\avg{\varepsilon_{m}}$ & 1636 & $4.6\times10^{-3}$ & --- & 5.7 \% & --- & 31 \%\\ % Magnetostriction
				\br
			\end{tabular}
			*Units: $M_s$ (A/M), $\lambda_s$ (ppm), $K$ (J/$\mathrm{m}^3$), $\beta$ ($\mathrm{m}^3$/J)\\ 
		\end{indented}
	\end{table}
	
	\begin{figure}[ht!]
		\centering
		\includegraphics[width = 15cm]{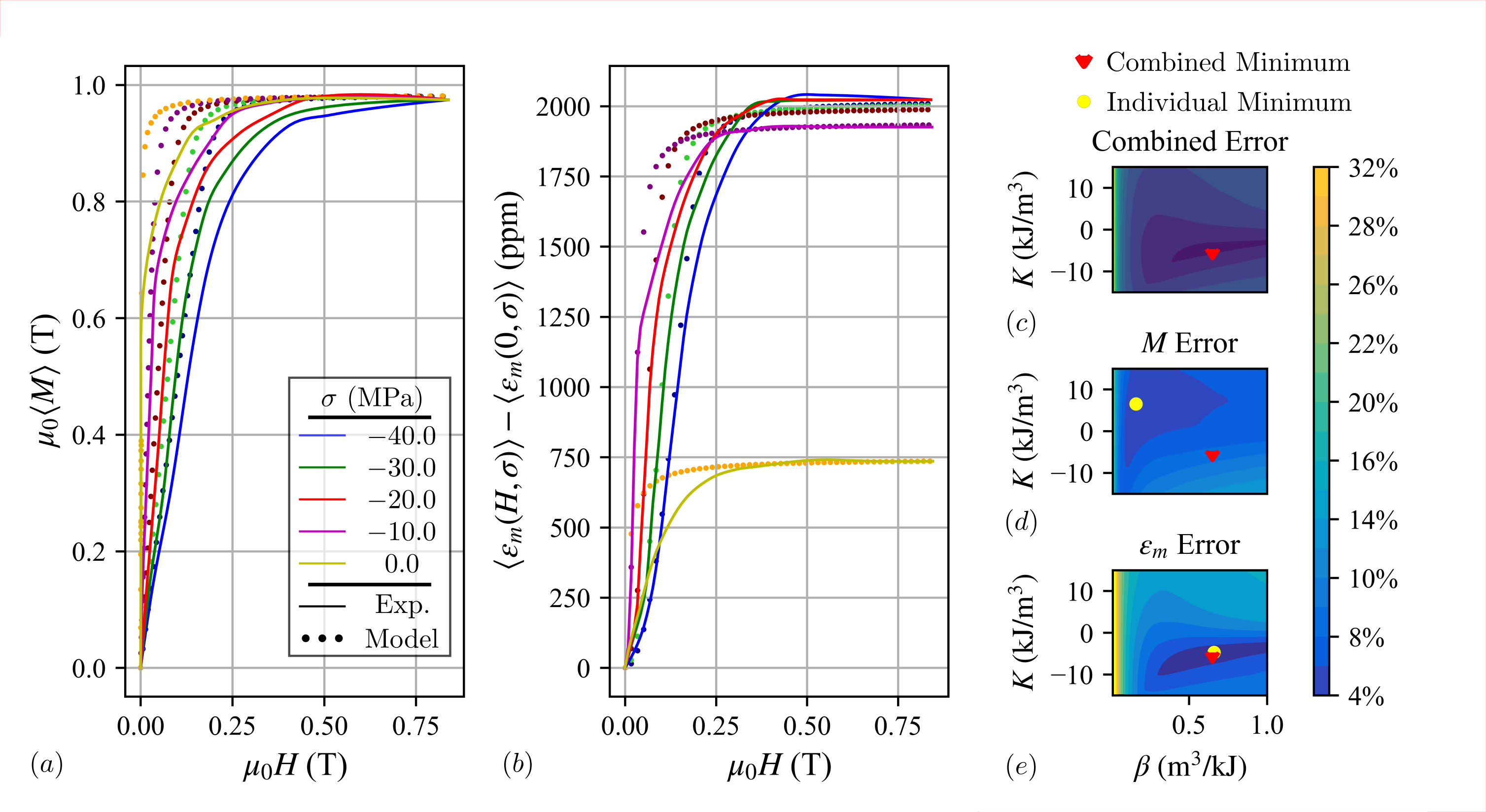}
		\caption{(a) Magnetization and (b) magnetostriction of $\textrm{Tb}_{0.3} \textrm{Dy}_{0.7} \textrm{Fe}_{19.2}$ at constant stress values compared to the 1D constitutive model with $\{K,\beta\}$ minimizing the combined error. Relative errors of (c) the combined data, (d) only the magnetization, and (e) only the magnetostriction for a parametric sweep of $K$ and $\beta$. In parts (c)-(e) the red triangular markers are the locations of minimum combined error, while the yellow circular markers are locations of of minimum magnetization or magnetostriction error. Data digitized from \cite{Zhao1998}. \label{fig:Terfenol}}
	\end{figure}
	Figure \ref{fig:Terfenol} parts (a) and (b) compare the model with experimental data for Terfenol-D using the combined fit parameters in Table \ref{table:properties}. The solid lines are the digitized experimental data, while the circular markers are modeled data points. Figures \ref{fig:Terfenol} (c)-(e) show the average relative error per data point for the data in \ref{fig:Terfenol}(a) and \ref{fig:Terfenol}(b). The red triangular markers show the location of the optimal values for the combined error, while the yellow circles are the locations of minimum error when fitting just the magnetization / magnetostriction. When using the $\{K,\beta\}$ that minimize the combined error, there  is $7.8\%$ and $3.0\%$ relative error per data point in the magnetization and magnetostriction comparison, respectively. It is worth noting that the best fit parameters for the magnetization \ref{fig:Terfenol}(d) and magnetostriction \ref{fig:Terfenol}(e) occur for different $\{K, \beta \}$. The combined result in \ref{fig:Terfenol}(c) is skewed towards the optimal magnetostriction location as a wide range of $\{K, \beta \}$ values produce errors close to the global magnetization minima of $5.0\%$ (as seen in \ref{fig:Terfenol}(d)). It is possible to reduce the average error to $5.0 \% $ if only the magnetization is compared, or $3.0 \%$ if only the magnetostriction is compared. 
	
\begin{figure}[ht!]
		\centering
		\includegraphics[width = 15cm]{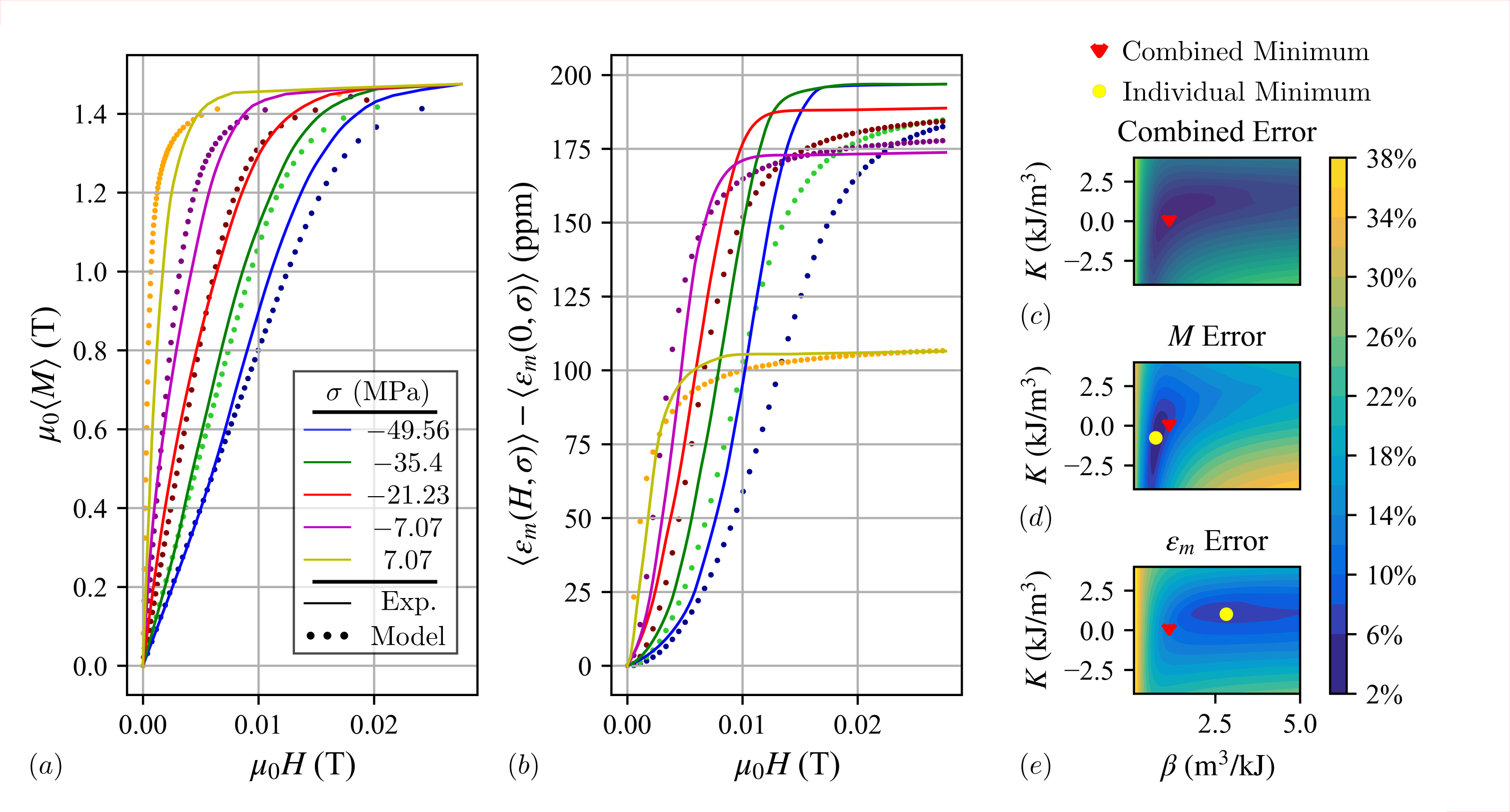}
		\caption{(a) Magnetization and (b) magnetostriction of $\textrm{Fe}_{79.1}\textrm{Ga}_{20.9}$ at constant stress values compared to constitutive model. Relative errors of (c) the magnetization (d) the magnetostriction and (e) the combination of the two over a parametric sweep of $K$ and $\beta$. In parts (c)-(e) the red triangular markers are the locations of minimum combined error, while the yellow circular markers are locations of of minimum magnetization or magnetostriction error. Data digitized from \cite{Mahadevan2010}. }
		\label{fig:Galfenol 1}
	\end{figure}

	Figure \ref{fig:Galfenol 1} parts (a) and (b) show a comparison of the experimental data for  $\mathrm{Fe}_{79.1} \mathrm{Ga}_{20.9}$ using the combined fit parameters, with formatting identical to \ref{fig:Terfenol}. Figures \ref{fig:Galfenol 1}(c)-(e) show the average relative error for the data in \ref{fig:Galfenol 1}(a) and \ref{fig:Galfenol 1}(b) is $6.4\%$ and $12\%$ per data point respectively. For this material the magnetization data is accurately described, while the model struggles to capture the magnetostriction data in compression. It is possible to reduce the average error to $4.0 \% $ if only the magnetization is compared, or $6.7 \%$ if only the magnetostriction is compared (i.e., the presented magnetization is close to its best fit, while the magnetostriction can be improved).
	
\begin{figure}[ht!]
		\centering
		\includegraphics[width = 15cm]{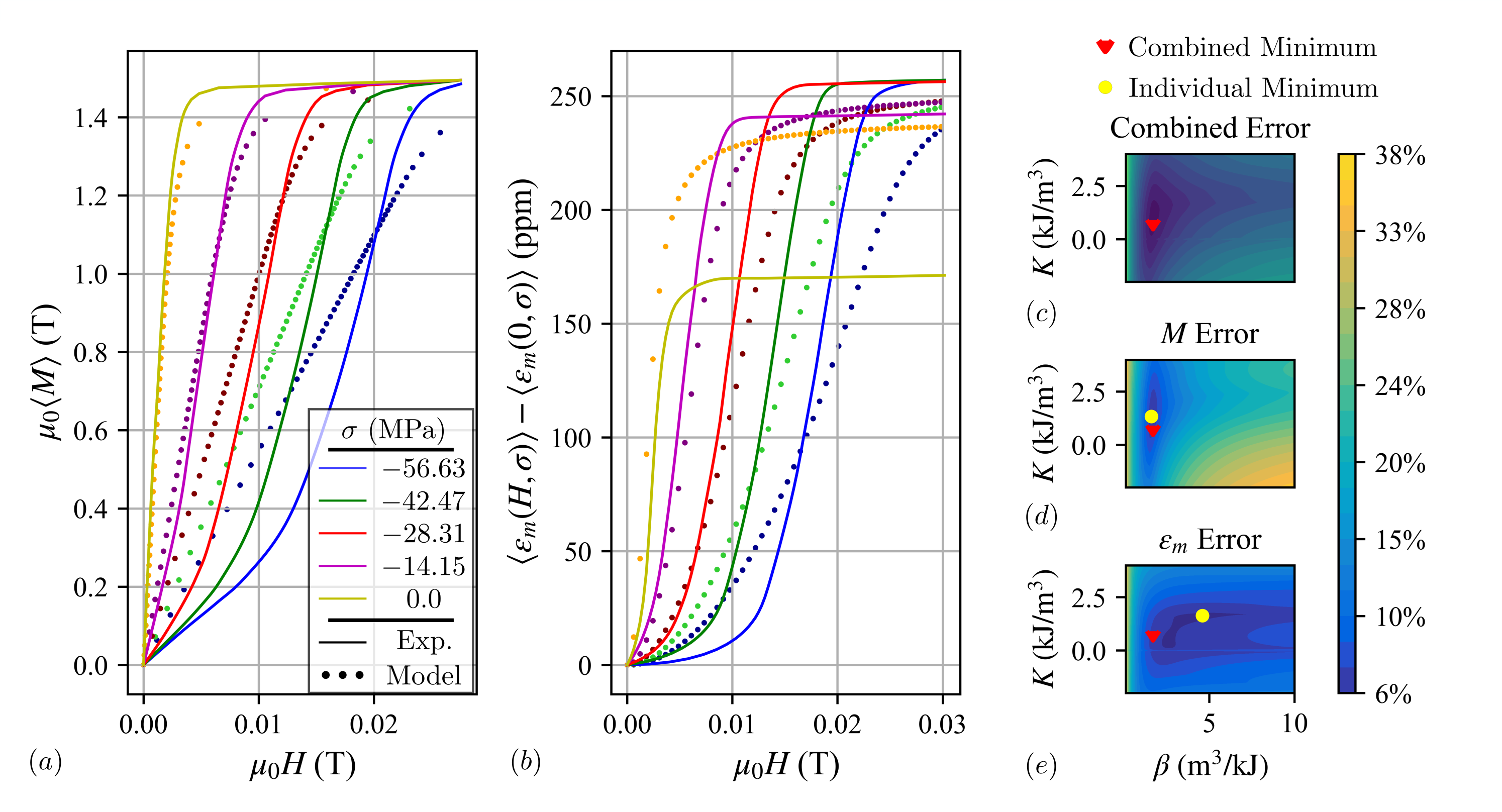}
		\caption{(a) Magnetization and (b) magnetostriction of $\textrm{Fe}_{81.6}\textrm{Ga}_{18.4}$ at constant stress values compared to constitutive model. Relative errors of (c) the magnetization (d) the magnetostriction and (e) the combination of the two over a parametric sweep of $K$ and $\beta$. In parts (c)-(e) the red triangular markers are the locations of minimum combined error, while the yellow circular markers are locations of of minimum magnetization or magnetostriction error. Data digitized from \cite{Mahadevan2010}.}
		\label{fig:Galfenol 2}
	\end{figure}

	Figure \ref{fig:Galfenol 2} parts (a) and (b) show a comparison of the experimental data for  $\textrm{Fe}_{81.6}\textrm{Ga}_{18.4}$ using the combined fit parameters. Figures \ref{fig:Galfenol 2} (c) shows the average relative error for the data in \ref{fig:Galfenol 2}(a) and (b) is $6.8\%$ and $7\%$ per data point respectively. Once again the combined result in \ref{fig:Galfenol 2}(c) is skewed towards the optimal magnetization. Despite this low relative error, a close look at \ref{fig:Galfenol 2}(a) shows that the constitutive model doesn't qualitatively capture the nonlinear (Type II) behavior that occurs when the material is subjected to large compressive stresses. As a result, we do not recommend using the analytical model for this composition of Galfenol. 
	
	Revisiting the anisotropy assumptions utilized in this model can help explain when the model is expected to accurately simulate experimental data. A relatively small restriction was placed on the isotropic magnetostrictive and Zeeman energies by assuming the principle stresses $\sigma_1 \neq \sigma_2 = \sigma_3$, with a magnetic field $\V{H} \parallel \hat{e}_1$. We note this mimics the field / stress combinations applied in the experimental studies compared to above, and therefore should not decrease the accuracy of the model. Even for a non-isotropic magnetostrictive material, as long as the stress and field are parallel to the $\avg{1 0 0}$ direction the model accurately describes the magnetostrictive energy. Conversely, a significant restriction was placed on the MCA by requiring it to be transversely isotropic. Due to this assumption the model is only capable of modeling Type I MH curves. The crystalline structure of Terfenol-D and Galfenol are typically cubic with MCA of the form, 
	\begin{align}
		f_{cubic} = K_1 (m_1^2 m_2^2 + m_2^2 m_3^2 + m_3^2 m_1^2 )
		+ K_2 (m_1^2 m_2^2 m_3^2).
	\end{align} 
	While the use of cubic MCA is expected to produce a more accurate model, there is no known closed form solution to the integral equations above once an MCA of this form is utilized. However, we can analyze the effect $K_1$ and $K_2$ have on magnetization curves by relying on numerical integration (e.g., Riemann sums), to compare cubic MCA to transversely isotropic. 
	
	\begin{figure}[ht!]
		\centering
		\includegraphics[width = 15cm]{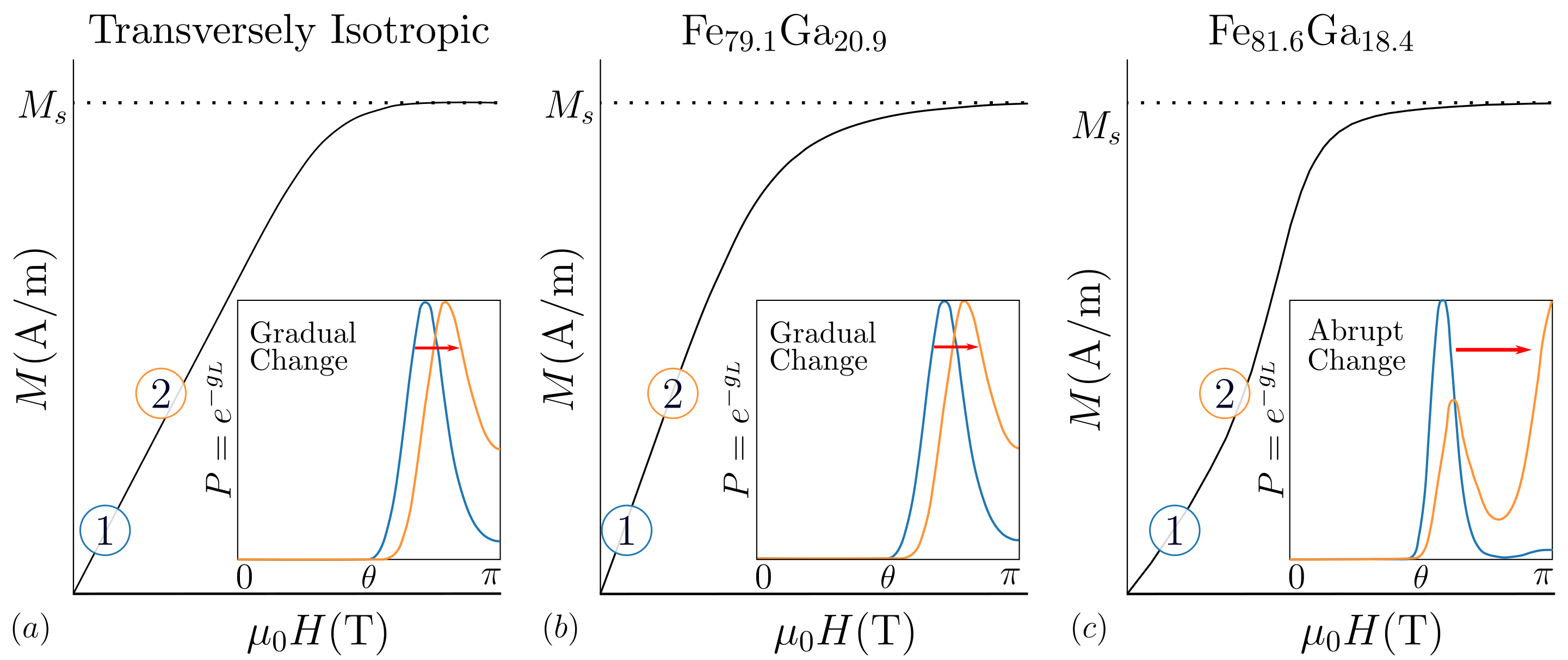}
		\caption{MH curves under $50$ MPa of compression with (a) Transversely Isotropic MCA, $K = -4500 \mathrm{J/m}^3$, (b) Cubic MCA for $\textrm{Fe}_{79.1}\textrm{Ga}_{20.9}$, $K_1 = -1e3 \mathrm{J/m}^3$ and $K_2=  1e4\mathrm{J/m}^3$, (c) and Cubic MCA for $\textrm{Fe}_{81.6}\textrm{Ga}_{18.4}$, $K_1 = 3.5e4$ $\mathrm{J/m}^3$ and $K_2 = -8e4$ $\mathrm{J/m}^3$ \cite{ Rafique2004}. Inset graphs contain the probability density of the systems at Points 1 and 2}
		\label{fig:Type_1}
	\end{figure}
	
	Figure \ref{fig:Type_1} (a) shows a Type I MH curve produced using the proposed model under $50$ MPa of compression, and a $K$ value of -4500 (J/$\mathrm{m}^3$), simulating the Terfenol-D modeled above. To understand how the MH curve saturates we examine the probability landscape of two points along the curve. For each point of interest, the probability density of the system in spherical coordinates is calculated using $P(\theta; H, \sigma, \beta) = \exp{(-\beta g_L)} / z$. The inset of Figure \ref{fig:Type_1} (a) plots the probability density against the possible magnetization directions $\theta$. Starting at Point 1, $\mu_0 H = 15$ mT in the negative $z$ direction, the applied field has moved the peak probability from $\theta = \pi/2$ a small amount towards $\theta = \pi$. As $\mu_0 H$ increases to $20$ mT and we transition along the MH curve to Point 2, the peak again gradually shifts to the right. As the field is continually increased past Point 2, the location of peak probability continues to gradually shift towards $\theta = \pi$. It is this gradual change in the probable direction of magnetization that produces Type I behavior. The only noticeable change in the MH curve is the final approach to saturation. This occurs when the location of peak probability is exactly at $\theta = \pi$, and instead of moving, it simply becomes more and more probable (i.e., corresponding to an increasingly large energy minima at that point).
	
	Figure \ref{fig:Type_1} (b) shows an MH curve for $\textrm{Fe}_{79.1}\textrm{Ga}_{20.9}$ produced using $K_1 = -1e3$ $\mathrm{J/m}^3$ and $K_2=  1e4$ $\mathrm{J/m}^3$ \cite{ Rafique2004} and placing the material under $50$ MPa of compression. For this MCA energy minima occur in the $\avg{1 0 0}$ family of directions when $h=0$ and $\sigma=0$. Applying stress causes the minima to appear exclusively in the $\avg{100}$, $\avg{\bar{1}00}$, $\avg{010}$, and $\avg{0\bar{1}0}$ directions (i.e., $\pm x$ and $\pm y$). To plot the 2D probability landscape in a single line the inset of Figure \ref{fig:Type_1} (b) shows $P(\theta) = \,P(\theta,\phi =\pi/4)$. In the inset of Figure \ref{fig:Type_1} (b) the exact same trend as the inset of Figure \ref{fig:Type_1} (b) is observed. Starting at Point 1, $\mu_0 H = 15$ mT, and moving to point 2, $\mu_0 H = 20$ mT, we see the same gradual shift in the probable direction of magnetization which produces Type I behavior. 
	
	Finally, \ref{fig:Type_1} (c) shows an MH curve for $\textrm{Fe}_{81.6}\textrm{Ga}_{18.4}$ produced using $K_1 = 3.5e4$ $\mathrm{J/m}^3$ and $K_2 = -8e4$ $\mathrm{J/m}^3$ \cite{ Rafique2004} and placing the material under $50$ MPa of compression. Of note the points of interest were specifically chosen to highlight the Type II transition from concave up to down and again occur at $\mu_0 H = 15$ mT and $20$ mT. Also as this is another cubic material, the single line plotted in the inset graph is the maximum probability, $P(\theta) = P(\theta,\phi =0)$. At Point 1 the direction with the highest probability is slightly to the right of $\theta = \pi/2$. However, as we move to Point 2 there is a significant change in the probability landscape. The global minima abruptly shifts to $\theta = \pi$ and an additional probability peak emerges. This abrupt change in the probable direction of magnetization is what causes the magnetization curve to change from concave up to down, and the presence of this second peak is what has kept the material from saturating. Due to the assumption of transversely isotropic MCA the model in this paper is incapable of simulating $\textrm{Fe}_{81.6}\textrm{Ga}_{18.4}$, or other materials with similar anisotropies that produce abrupt changes in the probability landscape. 
	
	\section{Conclusion}
	This paper provides an analytical one-dimensional constitutive model for magnetostriction. Closed form analytical solutions were provided to calculate the average magnetization, magnetostriction, susceptibility, compliance, and piezomagnetic coupling coefficient. Additionally, it was demonstrated that the analytical model maintains numerical accuracy over a large range of applied magnetic fields and stress / anisotropy conditions. Finally, the model was used to simulate experimental data for three different materials. This comparison only required fitting two model parameters to the data. Comparisons between the experimental and modeled results indicate that the model is capable of simulating Terfenol-D and is expected to also accurately describe certain cubic materials as long as they have probability landscapes that are well approximated as transversely isotropic. 
	
	\newpage
	\section{Appendix}
	
	\subsection{Low Anisotropy Series Expansion}
	\label{subsection: Appendix Series Expansion}
	
	As previously stated, the solutions become indeterminate when $A = 0$. By performing a series expansion of $\exp{(Am_1^2)}$ about $A=0$ we not only provide a solution for when $A=0$, but also obtain an accurate solution when $h>>|A|$. Expanding the exponential about $A=0$ and substituting it into the partition function we find that,
	\begin{align}
		z/2 \pi 
		&= \sum\limits_{n=0}^{N}  z_n 
		= \sum\limits_{n=0}^{N} \int\limits_{-1}^{1} \frac{(-A m_1^2)^n}{n!} 
		\exp{ (h m_1) } \dd m_1
		% Zn
	\end{align}
	which has the general solution
	
	\begin{equation}
		z_n 
		= 
		\frac{(2n)! (-A)^n (-h)^{-2n}}{h n!}
		\left( (-1)^{4n+1}  - 1 +
		\sum_{k=0}^{2n} \frac{(-h)^k\ee{h} + (-1)^{4n+1} h^k\ee{-h}}{k!} 
		\right). 
	\end{equation}
	This expression has been provided in terms of polynomial expansions, however slightly more compact expressions can be obtained in terms of gamma functions. As the expressions can become very lengthy, the authors utilized a computer algebra system (Mathematica) to simplify the expressions and export them for use in Matlab. 
	
	Using the thermodynamic relationships described above both the average material response and material properties can be found by taking partial derivatives of $z_n$ with respect to $h$ and $A$ for a desired level of accuracy, controlled by $N$. In the error analysis below, a value of $N=2$ was utilized.
	
	\subsection{Material Properties}
	\label{subsection:Appendix Material Properties}
	
	Magnetization and magnetostriction (Equations \eqref{eq:magnetization} and \eqref{eq:magnetostriction}) are proportional to $\avg{\V{m}}$ and $\avg{ \V{m} \tens \V{m} }$. When restricted to one dimensional behavior, this can be simplified to 
	\begin{align}
		\avg{ M }
		&= M_s \avg{m} \label{}\\
		\avg{\varepsilon_m} 
		&= \frac{3}{2} \lambda_s \avg{m^2} \label{}
	\end{align}
	Using this new notation the relationships in equations \eqref{eq:chi_general} - \eqref{eq:q_general}  can be rewritten as 
	\begin{align}
		\avg{\chi} 
		&= \beta \mu_0 M_s^2 
		\left[\, \avg{m^2} - \avg{m}^2 \,\right]
		\label{eq:chi_m}\\
		\avg{S_{m}} 
		&= \beta \left(\frac{3 \lambda_s}{2}\right)^2 
		\left[\, \avg{m^4} - \avg{m^2}^2 \, \right] 
		\label{eq:S_m}\\
		\avg{q} 
		&= \beta M_s \frac{3 \lambda_s}{2} 
		\left[\, \avg{m^3} - \avg{m}\avg{m^2}  \, \right] 
		\label{eq:q_m}
	\end{align}
	The definitions for $\avg{m}$ and $\avg{m^2}$ were previously presented in equations \eqref{eq:Magnetization 1} and \eqref{eq:Magnetostriction 1}. The only terms yet to be defined are the tensor products $\{\avg{ \V{m} \tens \V{m} \tens \V{m}},\, \avg{ \V{m} \tens \V{m} \tens  \V{m} \tens \V{m} } \}$ simplified in one dimension as $\{\avg{m^3}, \avg{m^4}\}$ respectively. These two terms and are shown here. 
	\begin{align}
		\avg{m^3}
		&= \begin{cases} 
			\frac{ 1 } { \tilde{z}_{+} 8 A^{7/2} } 
			\left( \sqrt{A} \left( 4 A^2 \left( 1 - e^{-2 h} \right)
			-2 A \left( e^{-2 h} (h - 2) + h + 2 \right)
			\right. \right. \\ \left.\left. 
			\qquad
			+ \left( 1 - e^{-2 h} \right) h^2 \right)
			\right)
			- \frac{6 A h + h^3}{8 A^3}
			&
			A > 0 \\
			\\
			\frac{ 1 } { \tilde{z}_{-} 8 A^3 } 
			\left( 4 A^2 \left( 1 - e^{-2 h} \right)
			-2 A \left( e^{-2 h} (h - 2) + h + 2\right)
			\right.  \\ \left.
			\qquad
			+ \left( 1 - e^{-2 h} \right) h^2 \right)
			+ \frac{ 6 A h - h^3 } { 8 A^3 }
			& 
			A < 0, \, \gamma_{+} > 0 \\
			\\
			\frac{ 1 } { \tilde{z}_{-} 8 A^3 }
			\left( e^{ -\text{$\gamma $1} ^ 2 - \text{$\gamma $2} ^ 2 } \left( \left( \frac{1}{2} 
			\left( e^h - e^{-h} \right) \left( 6 A h + 8 (A-1) A + 5 h^2 \right)
			\right. \right. \right. \\ \left. \left. \left.
			+ \frac{1}{2} \left( e^{-h} + e^h \right) h ( 2 A + 3 h) \right) 
			e^{ \text{$\gamma $1}^2 +h } - 3 e^{ \text{$\gamma $2}^2 } h ( 2 A + h ) \right) \right) 
			+ \frac{6 A h - h^3}{8 A^3}
			& 
			A < 0  ,\, \gamma_{+} < 0 
		\end{cases} \label{eq:m^3}
	\end{align}
	
	%m^4
	\begin{align}
		\avg{m^4}
		&= \begin{cases} 
			\frac{ 1 }{ \tilde{z}_{+} 8 A^4 } 
			\left( A \left( e^{-2 h} + 1 \right) \left( 4 A^2 - 6 A + h^2 \right) 
			\right. \\ \left. 
			- \frac{1}{2} \left( 1 - e^{-2 h} \right) h \left( 2 A (2 A-5) + h^2 \right) \right)
			+ \frac{12 A^2-12 A h^2+h^4}{16 A^4}
			&
			A > 0 \\
			\\
			\frac{ 1 }{ \tilde{z}_{-} 16 A^4} 
			\left( e^{-2 h} \left( 8 A^3 \left( e^{-2 h} + 1 \right) 
			-4 A^2 \left( e^{-2 h} (3 - h) + h + 3 \right) 
			\right. \right. \\ \left. \left. 
			+ 2 A h \left( e^{-2 h} (h - 5) + h + 5 \right)
			- \left( 1 - e^{-2 h} \right) h^3 \right) \right)
			+ \frac{12 A^2-12 A h^2+h^4}{16 A^4}
			& 
			A < 0, \, \gamma_{+} > 0 \\
			\\
			\frac{ 1 }{ \tilde{z}_{-} 16 A^4} 
			\left( e^{ -\text{$\gamma $2}^2 } \left( 8 A^3 \left( e^{2 h} + 1 \right) 
			+ 4 A^2 \left( h - e^{2 h} (h + 3) -3 \right)
			\right. \right. \\ \left. \left. 
			- 2 A h \left( -6 h e^{ \text{$\gamma $2}^2 - \text{$\gamma $1}^2 } 
			+ 5 e^{2 h} (h - 1) - h + 5 \right) 
			- h^3 \left( -6 e^{ \text{$\gamma $2}^2 - \text{$\gamma $1}^2 } 
			+ 7 e^{2 h} - 1 \right) \right) \right)
			\\
			+ \frac{12 A^2-12 A h^2+h^4}{16 A^4}
			&
			A < 0  ,\, \gamma_{+} < 0 
		\end{cases} \label{eq:m^4}
	\end{align}
	
	\subsection{Error Surfaces}
	\label{subsection: Appendix Error Surfaces}
	
	The following figures are the error surfaces generated by comparing the presented model to numerical integration. A grid of $N=100$ logarithmically spaced field points $10^{-2} \leq h\leq 10^{6}$ and $N=200$ logarithmically spaced anisotropies  $10^{-2} \leq |\pm A| \leq 10^{6} $ were generated for $N_{total} = 20,000$ points. At each grid point numerical integration was performed using Matlab's \texttt{integral()} function with relative and absolute errors of $10^{-12}$. The relative errors for each equation were calculated as $|f_{num} - f_{eqn}| / f_{num}$, where $f$ is the parameter of interest. 
	
\begin{figure}
		\centering
		\includegraphics{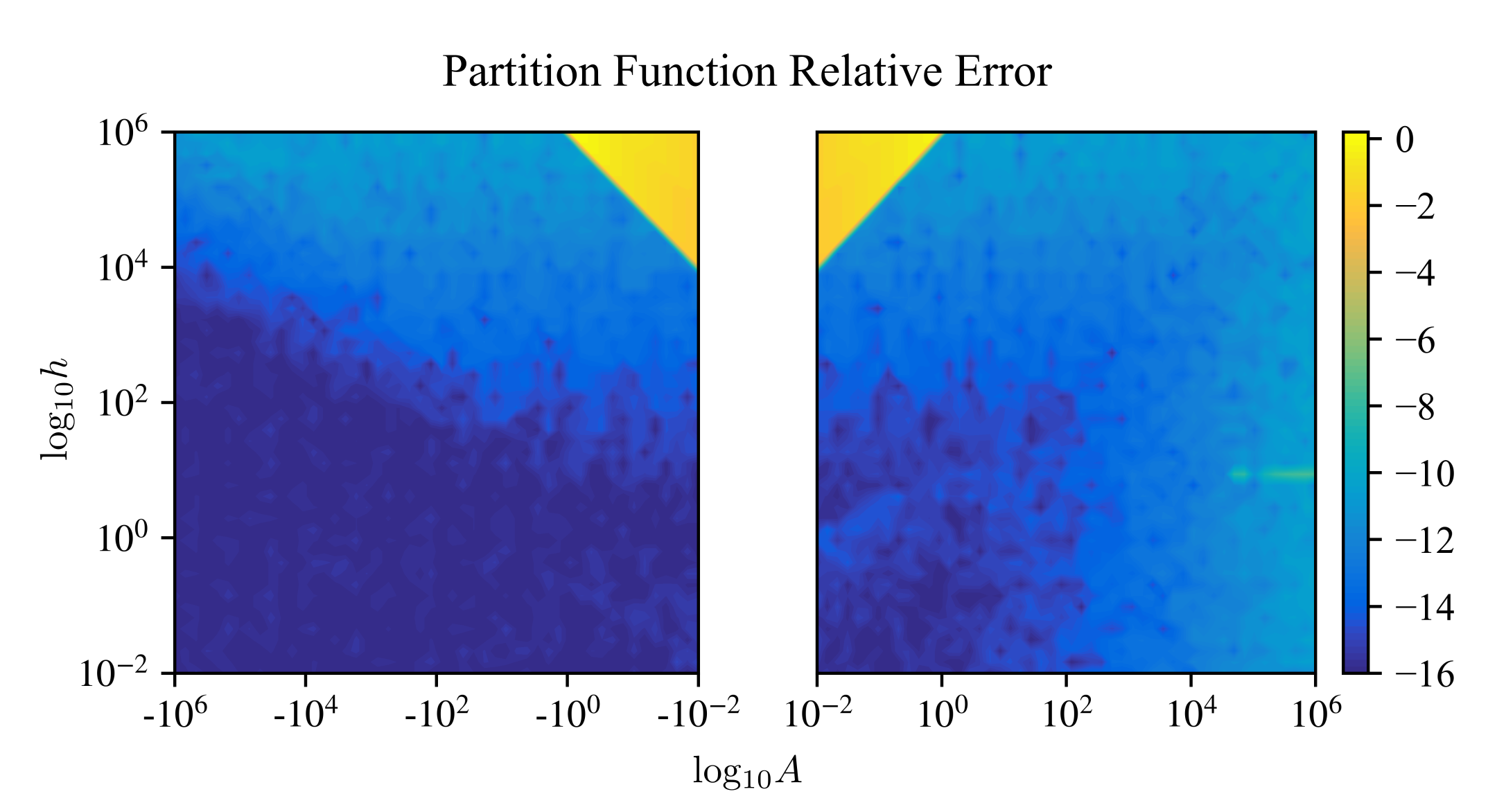}
		\caption{Logarithmically scaled relative error of the equation \eqref{eq:zprime} for $\tilde{z}$ compared to standard numerical integration with absolute errors  $10^{-12}$}
		\label{fig:Partition Error}
	\end{figure}

	Figure \ref{fig:Partition Error} examines equation \eqref{eq:zprime} for $\tilde{z}$ which notably includes the energy offsets shown in \eqref{eq:offsets}. Including these offsets changes the numerical value of the partition function, however only the slope of $\tilde{z}$ produces observable quantities, so the shift has negligible physical impact. The yellow triangular region, which has relative errors ranging from 10\% to 100\%, is where the low anisotropy series expansion was employed ($|A|/h < 10^{-7}$). While this error is quite high Figures \ref{fig:Magnetization Error} and \ref{fig:Magnetostriction Error} below show that the low anisotropy expansion maintained low relative errors for the magnetization and magnetostriction, respectively. Outside the expansion region the equations for $\tilde{z}$ maintains an average relative error near $10^{-14}$.  
	
	Figure \ref{fig:Magnetization Error} compares equation \eqref{eq:Magnetization 1} for $\avg{M}$ to numerical integration. Over all tested field values the relative error ranged from only $10^{-16}$ to $10^{-8}$ showing that the magnetization solutions maintain significant numerical accuracy for all applied fields and stresses. The error starts increasing as $h/|A|$ grows, however once the low anisotropy expansion is utilized the error returns to $~10^{-13}$. 
	
	Finally, Figure \ref{fig:Magnetostriction Error} compares equation \eqref{eq:Magnetostriction 1} for $\avg{\varepsilon_m}$ to numerical integration. The maximum observed error in this graph remains $\leq 10^{-3}$. While the error once again climbs as both $h/|A|$ grows, the low anisotropy expansion prevents the error from climbing any larger. 
	
	\begin{figure}
		\centering
		\includegraphics{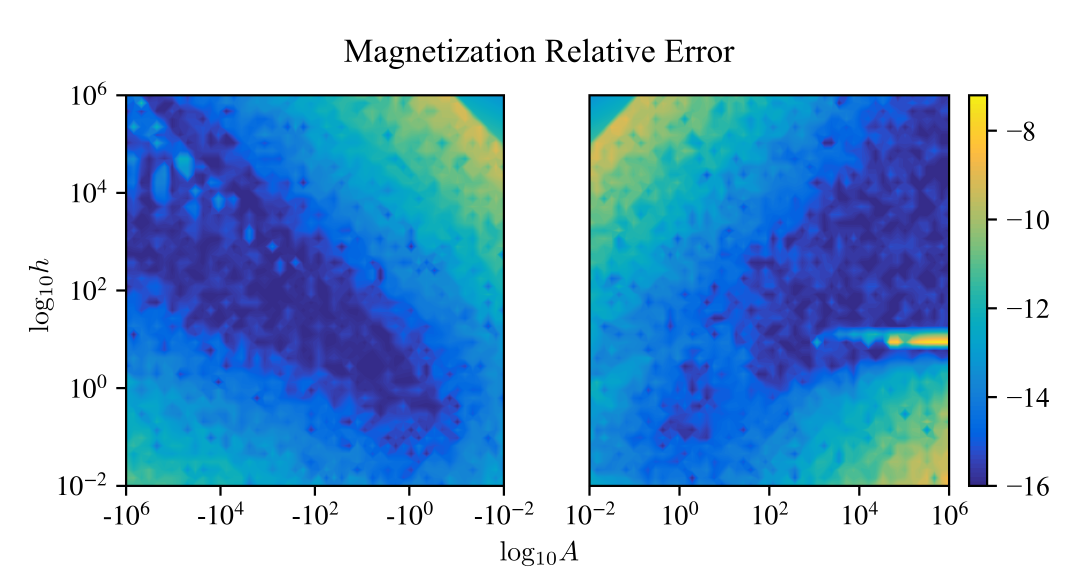}
		\caption{Logarithmically scaled relative error of equation \eqref{eq:Magnetization 1} for $\avg{M}$ compared to standard numerical integration with absolute errors  $10^{-12}$}
		\label{fig:Magnetization Error}
	\end{figure}
	
	\begin{figure}
		\centering
		\includegraphics{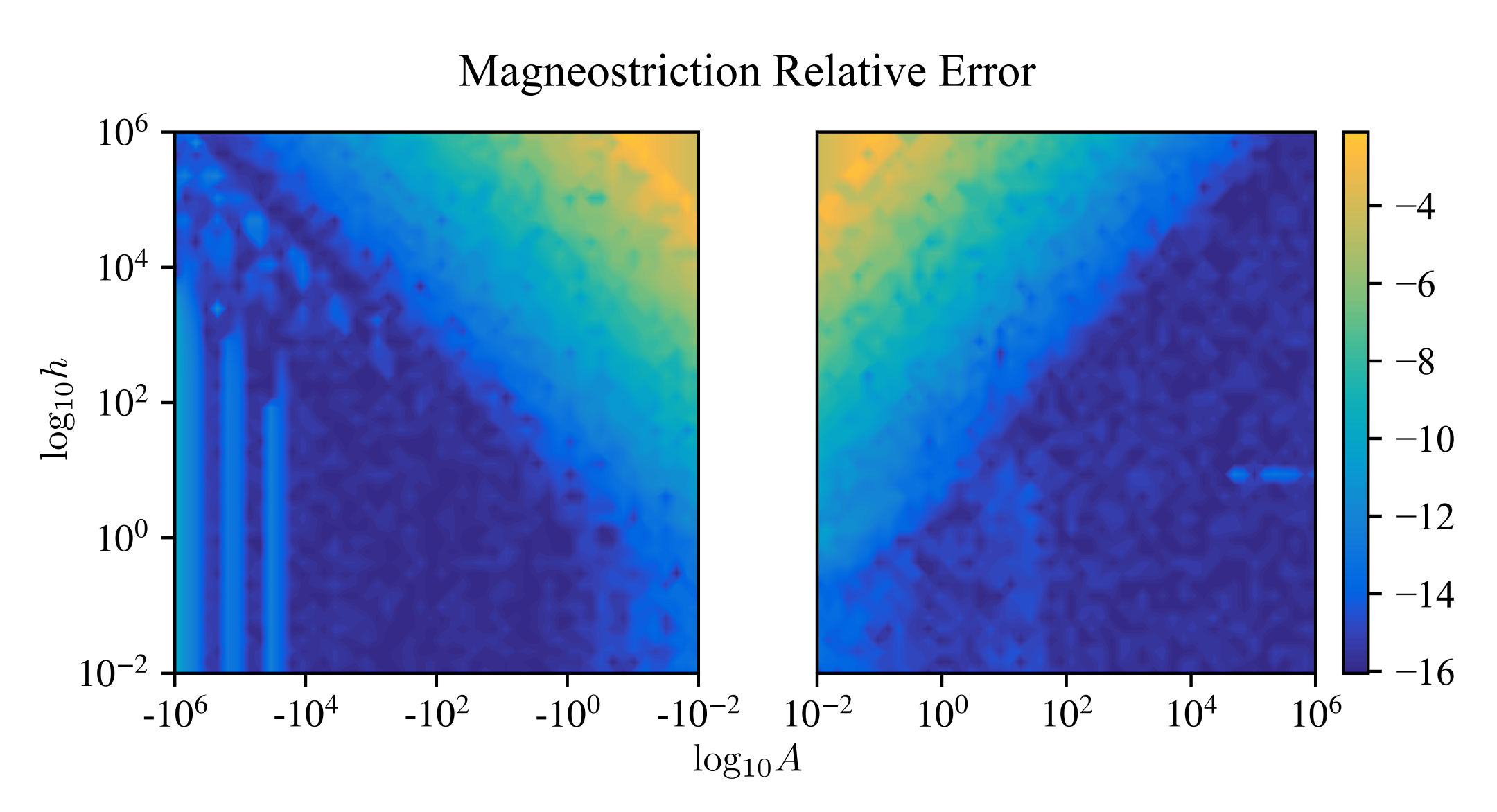}
		\caption{Logarithmically scaled relative error of equation \eqref{eq:Magnetostriction 1} for $\avg{\varepsilon_m}$  compared to standard numerical integration with absolute errors  $10^{-12}$}
		\label{fig:Magnetostriction Error}
	\end{figure}
	
	\newpage
	
	\section*{References}
	\bibliography{ref}
	
\end{document}